\documentclass[reprint,amsmath,amssymb,aps,twocolumn,superscriptaddress,prb]{revtex4-2}

\usepackage{upgreek}
\usepackage{graphicx}
\usepackage{dcolumn}
\usepackage{bm}
\usepackage[caption=false]{subfig}
\usepackage{float}
\usepackage{placeins}
\usepackage{xcolor}
\usepackage{siunitx}
\usepackage{titlesec}
\usepackage{braket}
\usepackage[colorlinks=True,linkcolor=red,citecolor=blue,urlcolor=blue]{hyperref}
\usepackage{lipsum}
\usepackage{enumerate}

\usepackage{booktabs}
\usepackage[ngerman,english]{babel}
\defineshorthand{"~}{\babelhyphen{nobreak}} 
\useshorthands{"}

\renewcommand{\vec}[1]{\mathbf{#1}}
\newcommand{\kdotpy}{\texttt{kdotpy}}
\newcommand{\kdotp}{\ensuremath{\mathbf{k}\cdot\mathbf{p}}}

\begin{document}
\preprint{V0.0}

\title{Self-Consistent Model for Gate Control of Narrow-, Broken-, and Inverted-Gap (Topological) Heterostructures}

\author{Maximilian Hofer$^{1,2}$, Christopher Fuchs$^{1,2}$, Moritz Siebert$^{1,2}$, Christian Berger$^{1,2}$, Lena Fürst$^{1,2}$, Martin Stehno$^{1,2}$,\\Steffen Schreyeck$^{1,2}$, Hartmut Buhmann$^{1,2}$, Tobias Kießling$^{1,2,\ast}$, Wouter Beugeling$^{1,2}$, and Laurens W. Molenkamp$^{1,2}$\\
\normalsize{$^{1}$Physikalisches Institut, Universität Würzburg, Am Hubland, 97074 Würzburg, Germany}\\
\normalsize{$^{2}$Institute for Topological Insulators, Am Hubland, 97074 Würzburg, Germany}\\
\normalsize{$^\ast$To whom correspondence should be addressed; E-mail: \texttt{tobias.kiessling@uni-wuerzburg.de}}
}

\date{\today}

\begin{abstract}
Even small electrostatic potentials can dramatically influence the band structure of narrow-, broken-, and inverted"~gap materials.
A quantitative understanding often necessitates a self"~consistent Hartree approach.
The valence and conduction band states strongly hybridize and/or cross in these systems.
This makes distinguishing between electrons and holes impossible and the assumption of a flat charge carrier distribution at the charge neutrality point hard to justify.
Consequently the wide-gap approach often fails in these systems.
An alternative is the full"~band envelope"~function approach by Andlauer and Vogl~\cite{Andlauer2009}, which has been implemented into the open-source software package \kdotpy{} \cite{Beugeling2025}.
We show that this approach and implementation gives numerically stable and quantitatively accurate results where the conventional method fails by modeling the experimental subband density evolution with top"~gate voltage in thick (\qtyrange[range-phrase=~--~]{26}{110}{nm}), topologically inverted HgTe quantum wells.
We expect our openly-available implementation to greatly benefit the investigation of narrow-, broken-, and inverted"~gap materials.
\end{abstract}

\maketitle

\section{Introduction}

Exploration of novel physical effects in narrow-, broken-, and inverted"~gap materials like the search for Majoranas \cite{Ren2019,Liu2025} or tunable topological band inversion \cite{Li2009,Yang2008,Meyer2024,Vezzosi2025}, and subsequent realization of functional devices \cite{Nauschuetz2023,UamanSvetikova2023,Svetikova2024,DaguaConda2025}, requires precise band structure modeling.
Accurate modeling is particularly challenging in these systems, as small external or built"~in electric fields can dramatically change the band structure \cite{Bruene2014,Ziegler2020,Mahler2021,Wang2024,PfeufferJeschke1998,Becker2002,Becker2004a,Novik2005,Minkov2022,Zhang2001,Andlauer2009,Gospodaric2020,Li2009,Yang2008,Meyer2024,Nichele2017,Vezzosi2025}.

Empirical toy-potential models \cite{Bruene2014,Ziegler2020,Mahler2021,Wang2024} often describe experiments only qualitatively.
Instead, the Schrödinger and Poisson equation need to be iteratively solved self"~consistently to obtain a quantitatively accurate electrostatic Hartree potential and underlying band structure \cite{PfeufferJeschke1998,Becker2002,Becker2004a,Novik2005,Minkov2022,Zhang2001,Vezzosi2025}.
The iterative method requires extracting the charge density $\rho(z)$ along the growth direction of the heterostructure $z$ from the band structure.
The conventional (wide"~gap) approach assumes that the occupied states can be unambiguously separated into electron- and hole"~like, and that $\rho_0(z)\equiv 0$ at the charge neutrality point (CNP) \footnote{The CNP is defined as the chemical potential for which the system is charge neutral \emph{on average}, $n = \int \rho(z)dz=0$.} \cite{PfeufferJeschke1998,Becker2002,Becker2004a,Novik2005,Minkov2022,Zhang2001}.
In broken- or narrow"~gap heterostructures subject to strong electrostatic potentials, the valence and conduction band states hybridize strongly and/or cross over, violating these assumptions.
In practice, numerical instabilities, like failure to reach convergence occur.
This prohibits using the conventional approach for thicker devices (e.g., $\gtrsim \SI{30}{nm}$ for HgTe).
An alternative is given by the full"~band envelope"~function approach (FB"~EFA) \cite{Andlauer2009} which, similar to atomistic approaches, treats all states as a single carrier type and avoids these problems.

In this work, we test the implementation of the FB"~EFA, included in the open-source software package \kdotpy{}, developed by some of us \cite{Beugeling2025}.
We simulate the Hartree potentials and band structures of thick (\qtyrange[range-phrase=~--~]{26}{110}{nm}) HgTe quantum well heterostructures under tensile strain, a realization of the semimetallic three-dimensional topological insulator (3DTI) phase of the material.
The effect of single"~sided electrostatic gating is modeled by suitable boundary conditions.
We find excellent quantitative agreement of the simulated subband densities with the frequencies of the Shubnikov"~de~Haas oscillations in the transverse conductivity $\sigma_{xx}$ in magnetotransport experiments.


\section{Modeling}
\label{sec:modeling}

\subsection{Wide-Gap Approach vs. FB-EFA}

\begin{figure}
 \includegraphics[width=\columnwidth]{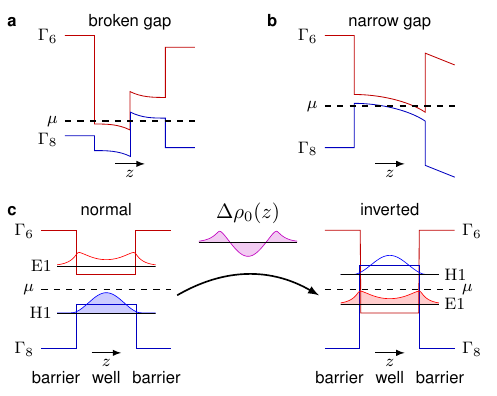}
 \caption{Illustration of the conceptual problems of the conventional approach.
 (a) In a heterostructure of broken-gap type, the valence (blue) and conduction (red) band edges overlap and hybridize strongly, and charge transfer can occur.
 (b) In a narrow-gap heterostructure, electric potentials can strongly bend the valence and conduction band edges, also leading to mixing and hybridization.
 In both cases, an unambiguous separation into electron- and hole-type states is not possible. 
 (c) In a thin HgTe quantum well with normal band ordering (left), the H1 subband is occupied if the chemical potential $\mu$ lies at the CNP, and the E1 subband is empty (wave functions $|\psi(z)|^2$ shown as red and blue curves, respectively).
 For an inverted quantum well (right), the subband order is reversed.
 It follows that the carrier distribution $\rho_0(z)$ at the CNP differs by $\Delta\rho_0(z)\propto |\psi_{\mathrm{E}1}(z)|^2 - |\psi_{\mathrm{H}1}(z)|^2$, the functional dependence of which is shown in the center of the figure.
 }
 \label{fig:cartoon}
\end{figure}

In each iteration step of the self-consistent Hartree method, the charge density profile $\rho(z)$ at the chemical potential $\mu$ needs to be calculated.
It is in this step that problems can arise in the conventional wide-gap approach.
In both the conventional method and the FB"~EFA, $\rho(z)$ is calculated by integrating the probability densities $|\psi_j(\vec{k}, z)|^2$ of all states $j$ \cite{Novik2005,Beugeling2025,Andlauer2009}:
\begin{equation}\label{eq:rhoz}
  \rho(z) = \rho_0(z) + \sum_j\frac{-e}{(2\pi)^2}\int |\psi_j(\vec{k}, z)|^2 f(E_j(\vec{k})-\mu) d^2\vec{k}\,,
\end{equation}
where $f(E)$ is a signed occupation function.
The discrete solutions of the Hamiltonian at each $k$ point are grouped into a set of subbands based on their relative energetic position, such that no band crossings occur \footnote{{When evaluating the subband densities, we group states into bands based on maximizing the wave function overlap between neighboring $k$ points. This allows band crossings, which are important to take into consideration in strong electrostatic potentials.}}.
In the wide"~gap approach, all states of a subband are treated as either holes ($f(E)<0$) or electrons ($f(E)>0$), based on the subband position relative to the CNP at $k=0$ \cite{Beugeling2025}.
This clear separation into electron- and hole"~like states is often impossible in narrow-, broken- \cite{Andlauer2009}, or inverted"~gap systems (such as thick HgTe quantum wells), and/or in strong electric potentials (compare Figure~\ref{fig:cartoon}a,b).

Moreover, the assumption from the wide-gap approach that $\rho_0(z) \equiv 0$ at the CNP  both for systems with normal and inverted band ordering leads to a conceptual problem:
For a normally ordered quantum well at the CNP, the heavy-hole subband H1 is occupied while the electronic subband E1 is empty.
In an inverted quantum well, the situation is reversed.
Thus, one expects the difference in $\rho_0(z)$ to be proportional to $|\psi_{\mathrm{E}1}(z)|^2 - |\psi_{\mathrm{H}1}(z)|^2$, as illustrated in Figure~\ref{fig:cartoon}c.
We conclude that the assumption $\rho_0(z) \equiv 0$ must be violated for at least one of these cases.

The FB"~EFA avoids both issues by filling all states with a single carrier type \cite{Andlauer2009}.
A suitable uniform background density ascertains that $\int\rho(z)dz = 0$ at the CNP, instead of postulating that $\rho_0(z)\equiv 0$ as in the wide-gap approach.

\subsection{Heterostructures}

We model a \SI{45}{nm} thick HgTe quantum well, encapsulated by $\sim\SI{50}{nm}$ thick Hg$_{0.3}$Cd$_{0.7}$Te barrier layers, pseudomorphically strained to a CdTe substrate.
The top-side electrostatic gate consists of a \SI{15}{nm} thick dielectric layer of HfO$_x$, followed by a \SI{2}{nm} Ti sticking"~layer and \qtyrange[range-phrase=~--~]{70}{100}{nm} Au gate electrode.
Analyses of three additional \SI{26}{nm}, \SI{70}{nm}, and \SI{107}{nm} thick quantum wells are presented in the Supporting Information.

\subsection{Single-Sided Gating}

We seek self-consistent solutions for the Hartree potentials and band structures at different gate voltages by employing the FB"~EFA of \kdotpy{} \cite{Beugeling2025} as described above.
It involves full diagonalization of the conduction band (which has fewer states than the valence band) and uses the top of the spectrum as the reference.
We calculate the band structure of the (Hg,Cd)Te/HgTe/(Hg,Cd)Te stack.
The influence from the electrostatic gate is modeled by suitable boundary conditions.
In the next section, we expand the model to the HfO$_x$ gate dielectric and Ti/Au gate electrode.
We use the eight-orbital Kane model \cite{Kane1957,Novik2005}, implemented in \kdotpy{} \cite{Beugeling2025}, discretized along the growth direction $z$ of the structure.
Terms with non"~axial symmetry and bulk"~inversion asymmetry (BIA) are included.
Further implementation specific details are discussed in Ref.~\cite{Beugeling2025}.

The material parameters are taken from Refs.~\cite{Novik2005,Beugeling2025}.
Following recent spectroscopic evidence, we adjust the squared Kane matrix element to $E_\mathrm{P}=\SI{20.8}{meV}$, and the valence band offset between HgTe and CdTe to $E_\mathrm{VBO}=\SI{-620}{meV}$ \cite{Zholudev2012,Bovkun2025}.
We are aware of the ongoing debate about the dielectric constant of HgTe possibly differing from its established value \cite{Ziegler2020,Bruene2014,Kozlov2016,Mahler2021,Wang2024}.
Here, we use the established values $\epsilon_\text{HgTe}=20.8$ and $\epsilon_\text{(Hg,Cd)Te}=13.6$ \cite{Baars1972,Novik2005,Beugeling2025}.

The eigenstates $\psi_j(\vec{k}, z)$ in eq.~\ref{eq:rhoz} follow from the Schrödinger equation
\begin{equation}\label{eq:schrodinger}
 \hat{H}(\vec{k})\ket{\psi_j(\vec{k})} = E_j(\vec{k})\ket{\psi_j(\vec{k})}\,
\end{equation}
where the Hamiltonian $\hat{H}$ is the sum of the Kane Hamiltonian and the Hartree potential $U(z)$.
The Hartree potential [$U(z)=-eV(z)$, in terms of the electric potential $V(z)$] and the charge density $\rho(z)$ are related by the Poisson equation,
\begin{equation} \label{eqn_poisson}
	\partial_z \left[\epsilon(z)\,\partial_z U(z)\right] = \frac{e}{\epsilon_0}\rho(z)\,,
\end{equation}
where $\epsilon(z)$ is a piecewise constant function with the dielectric constants of the layers.

To find suitable boundary conditions for a top-gated sample with finite doping density, we employ physical intuition:
We expect the charge screening length to be shorter than the sample thickness.
Due to the semimetallic band structure of thick HgTe quantum wells, this also applies in depletion mode, as the remaining charge carriers are able to screen the electric field efficiently.
Hence we set the electric field in the bottom barrier constant but non"~zero.
A more detailed justification for this choice of boundary conditions and procedure for calculating the value of the finite electric field in the bottom barrier is given below.

In Figure~\ref{fig:potential_45nm} we present the Hartree potentials obtained from the self-consistent calculations, spanning the experimentally probed density range.
While the electric field in the bottom barrier is fixed, the different densities naturally result in different electric fields inside the top barrier, which are experimentally defined by the gate electrode.
The potential drop across the HgTe layer is strongly nonuniform and the screening length for the electric field from the top barrier depends heavily on the total carrier density.
This hints at a complex charge carrier distribution.
Figure~\ref{fig:dispersion_45nm} shows the band dispersions at three representative densities.
The quantum wells are in the 3DTI semimetallic phase, which has two topological surface states (TSS), localized at the top and bottom quantum well interface, traversing the bandgap between the first bulk valence (VB) and conduction subbands (CB) \cite{Bovkun2025,Fuchs2025,Minkov2022,Gospodaric2020}.
Figure~\ref{fig:dispersion_45nm}b shows the dispersion for a near-symmetric Hartree potential.
Figure~\ref{fig:dispersion_45nm}a shows the dispersion at a smaller (p"~type) and Figure~\ref{fig:dispersion_45nm}c at a larger density.
The asymmetric Hartree potentials in a,c cause a strong splitting between the initially degenerate bands in b, which is especially pronounced for the VB and TSS.

\begin{figure}
    \centering
    \includegraphics[]{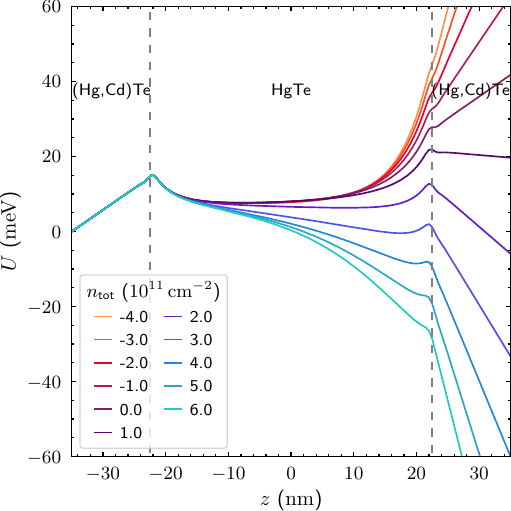}
    \caption{Self-consistently calculated Hartree potentials of a \SI{45}{nm} thick HgTe quantum well for several total carrier densities $n_\text{tot}$.
    The potentials $U$ are plotted along the growth direction $z$ of the quantum well.
    The HgTe well region is indicated by two dashed lines.}
    \label{fig:potential_45nm}
\end{figure}

\begin{figure*}[t]
    \centering
    \includegraphics[]{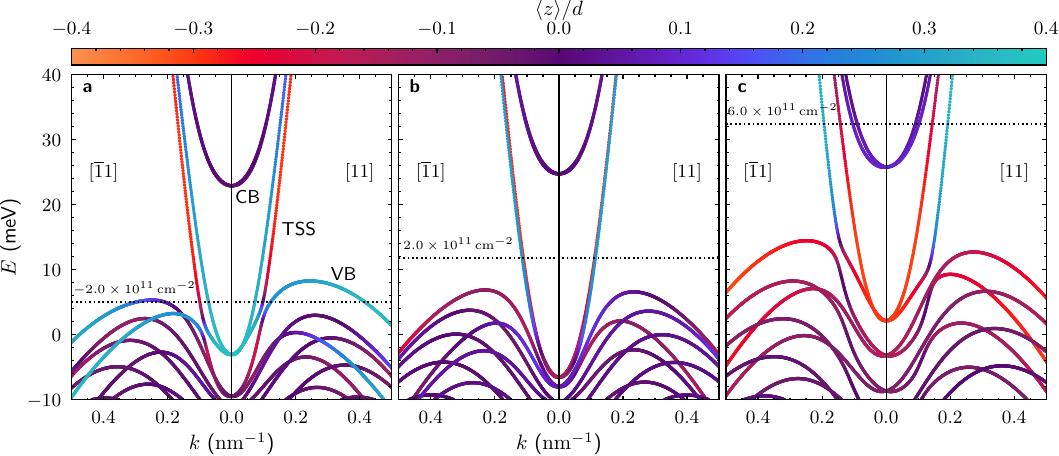}
    \caption{Energy dispersions $E$ of a $d=\SI{45}{nm}$ thick HgTe quantum well along the in"~plane $k$"~directions $[\overline{1} 1]$ and $[1 1]$ from three self-consistent Hartree \kdotp{} calculations shown in Figure~\ref{fig:potential_45nm}.
    (a) For a total carrier density of $n_\text{tot}=\SI{-2.0e11}{cm^{-2}}$, (b) $n_\text{tot}=\SI{2.0e11}{cm^{-2}}$, resulting in a near"~symmetric Hartree potential for the chosen boundary conditions (cf.~Figure~\ref{fig:potential_45nm}), and (c) for $n_\text{tot}=\SI{6.0e11}{cm^{-2}}$.
    The dashed lines give the position of the chemical potential.
    Labels in (a) indicate the topological surface states (TSS) and the first bulk valence (VB) and conduction (CB) subbands.
    The color coding shows the spatial expectation value $\langle z \rangle$ of the wave functions along the quantum well growth direction $z$, normalized by the quantum well thickness $d$.}
    \label{fig:dispersion_45nm}
\end{figure*}

\subsection{Subband Densities}

The experiment probes the occupation of the TSS, CB, and VB.
While it is necessary to employ the FB"~EFA for calculating the $z$"~resolved carrier density distribution during the self-consistent iteration, the total density in these bands is well defined (essentially by their Fermi vector), and the traditional method can be employed for extracting the individual subband densities.
Following the mass"~action law, the TSS and CB states are counted as n"~type carriers, while the VB states are counted as p"~type \cite{Beugeling2025}.
The subband densities are evaluated by numerical integration of the band structure in $k$ space.
It is sufficient to include states only up to $|k|=\SI{1}{nm^{-1}}$ (around \SI{15}{\percent} of the Brillouin zone of HgTe, cf. Ref.~\cite{Andlauer2009}), as the calculated subband densities no longer change significantly beyond this point.
While the TSS and CB states are almost axially symmetric, the VB states are strongly anisotropic in reciprocal space.
Due to combined breaking of the structure and bulk inversion symmetry, the hole pockets along the $[11]$/$[\overline{1}\overline{1}]$ and $[1\overline{1}]$/$[\overline{1}1]$ axes are not equivalent (see Figure~\ref{fig:dispersion_45nm} and Supporting Information).
The remaining symmetry allows limiting the angular range to a quarter"~circle centered at the $\Gamma$ point.
Thermal broadening of the Fermi"~Dirac distribution $f(E)$ is neglected as the experiments were performed at $<\SI{40}{mK}$.

\subsection{Gate Voltage}

For a complete electrostatic model of the experiment, the gate voltage can also be calculated.
The Hartree potential at the end of the top barrier $U(z_\text{top})$ (cf.~Figure~\ref{fig:potential_45nm}) is continued by treating the gate dielectric and gate electrode as one side of a parallel plate capacitor.
In the experiment, the voltage is applied between the quantum well (which is at the chemical potential $\mu$) and the gate electrode.
An additional offset term $\Delta\mu$ is introduced to account for the work function offset between the gate metal and barrier, which needs to be experimentally determined.
Thus, the gate voltage $V_g$ satisfies the relation
\begin{equation}
	eV_g = - \left[U(z_\text{top}) - \mu - \Delta\mu\right] - \frac{\epsilon_\text{(Hg,Cd)Te}}{\epsilon_\text{HfO$_x$}}d_\text{HfO$_x$}U'(z_\text{top})\,. \label{eq:VG}
\end{equation}
The potential $U(z)$ and its spatial derivative $U'(z)$ are evaluated inside the top barrier at the interface to the gate dielectric. 
$\epsilon_\text{(Hg,Cd)Te}$ and $\epsilon_\text{HfO$_x$}$ are the dielectric constants of the barrier and the HfO$_x$ gate dielectric respectively, and $d_\text{HfO$_x$}$ is the thickness of the gate dielectric.
The dependence on the chemical potential $\mu$ takes the quantum capacitance effect into account \cite{Luryi1988}.

\section{Magnetotransport and Data Analysis}
\label{sec:magnetotransport_analysis}

For transport measurements, standard \mbox{($600 \times 200)\,\si{\mu m^2}$} Hall bars are fabricated from the MBE grown structures \cite{Bendias2018,Shekhar2022}.
Magnetotransport measurements are performed at $<\SI{40}{mK}$ using standard low"~frequency lock"~in techniques.
A set of transport results from the same samples has already been analyzed in \cite{Fuchs2025}.

Measurement results of the longitudinal resistance $R_{xx}$ at four representative gate voltages are presented in Figure~\ref{fig:FFT_45nm}a.
In case of a single band with carrier density $n$, $R_{xx}$ oscillates with frequency $f_{B^{-1}} = h/(ne)$, when plotted against reciprocal magnetic field $B^{-1}$ \cite{Hinz2006,Nichele2017}.
It is easy to see in the raw data that more than one oscillation period is present.
The \SI{0.0}{V} curve, e.g., appears to show two distinct oscillation periods, indicating contributions from two occupied subbands, while the other traces show even more distinct oscillation periods.

In the following, we perform a fast Fourier transform (FFT) to extract the frequencies of the Shubnikov"~de~Haas oscillations pertaining to individual subbands.
First, we calculate the longitudinal conductivity $\sigma_{xx}$ and interpolate and re"~sample it to be linearly spaced in $1/B$.
Next, a linear background is subtracted and the resulting array is windowed using a Hamming window function and then FFT'd.
The reciprocal magnetic field frequencies $f_{B^{-1}}$ are converted to carrier densities $n$ using $n=(e/h)f_{B^{-1}}$.
This procedure is performed for all measured gate voltages.
The individual curves are then assembled to an FFT chart that shows the magnitude of the FFT curves (power spectral density, PSD) as a function of carrier density $n$ and gate voltage $V_g$ (see Figure~\ref{fig:FFT_45nm}b).
We see a clear evolution of the carrier densities (inverse frequencies) with gate voltage.
We also plot the experimental evolution of the total carrier density $n_\text{tot}(V_g)$ as a function of gate voltage (black line), extrapolated from low"~field Hall effect measurements at high n"~type densities \cite{Fuchs2025}.

\section{Discussion}
\label{sec:discussion}

\begin{figure}[h!]
    \centering
    \includegraphics[]{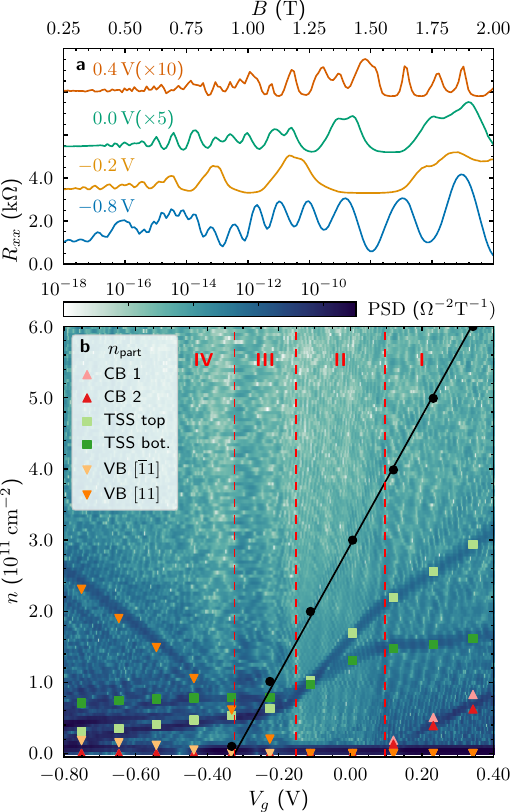}
    \caption{(a) Experimental longitudinal resistance curves $R_{xx}$ in an out"~of"~plane magnetic field $B$ at different gate voltages of a \SI{45}{nm} thick HgTe quantum well.
    The absolute scale applies to the \SI{-0.8}{V} curve.
    For clarity, the other curves are offset vertically and the \SI{0.0}{V} and \SI{0.4}{V} curves are scaled as indicated.
    (b) Extracted subband densities.
    Colored symbols show the individual subband densities $n_\text{part}$ of both topological surface states (TSS), the first bulk conduction bands (CB), and the hole pockets at the top of the valence band (VB) from self"~consistent Hartree \kdotp{} calculations (cf.~Figs. \ref{fig:potential_45nm} and \ref{fig:dispersion_45nm}).
    The background shows the power spectral density (PSD) at density $n$ obtained from FFTs of the experimental low-field Shubnikov"~de~Haas oscillations at different gate voltages $V_g$ on a logarithmic scale.
    The black line corresponds to the experimental gate action for the total carrier density obtained from Hall measurements.
    The black dots give the total carrier density from the calculations.
    Red"~dashed lines divide the gate voltage range into four transport regimes I-IV, see discussion and Ref.~\cite{Fuchs2025}.
    }
    \label{fig:FFT_45nm}
\end{figure}

We now seek to compare the data in Figure~\ref{fig:FFT_45nm} with the carrier densities from our band structure calculations.
For calculating the gate voltages, we treat $\Delta\mu$ and $\epsilon_\text{HfO$_x$}$ in eq.~\ref{eq:VG} as free parameters.
We adjust these parameters to best match the calculated total carrier density points $n_\text{tot}(V_g)$ to the experimental curve (black line in Figure~\ref{fig:FFT_45nm}b).
The total density in the calculation (black dots in Figure~\ref{fig:FFT_45nm}b) is given by the sum of the subband densities (taking into account the valley degeneracy for the VB states).
After adjusting the parameters in eq.~\ref{eq:VG}, the simulated gate voltages from the simple capacitor model agree well with the linear experimental gate action.
We discuss the obtained values for $\Delta\mu$ and $\epsilon_\text{HfO$_x$}$ across all four samples in the Supporting Information.

The calculated subband densities (colored symbols in Figure~\ref{fig:FFT_45nm}) as a function of gate voltage are in excellent agreement with the experimentally observed density signatures (dark background lines).
The splitting between the top and bottom TSS is captured very well by the calculation.
In the thin quantum well limit, this effect is the well known gate"~controllable, giant Rashba"~Bychkov"~type splitting in HgTe quantum wells \cite{Winkler2003,Minkov2022,Novik2005,Hinz2006}.
It naturally arises from the different gate action of the top"~gate on the top and bottom TSS due to electrostatic screening by the carriers in the different subbands.
The gate voltage range can be divided into four regimes I"~IV (marked in Figure~\ref{fig:FFT_45nm}b), depending on which carriers contribute to the electronic transport \cite{Fuchs2025}.
In regime II, only the TSS are occupied (see Figure~\ref{fig:dispersion_45nm}b for a corresponding dispersion plot).
This explains the still relatively large gate action of the top"~gate on the bottom TSS.
Only the top TSS partially screens the electric field and the potential over the entire HgTe well region changes quite strongly (cf. Figure~\ref{fig:potential_45nm} for $n_\text{tot}\sim\SI{1.6e11}{cm^{-2}} - \SI{3.8e11}{cm^{-2}}$).
In regime I, bulk CB states are occupied as well (compare Figure~\ref{fig:dispersion_45nm}c), leading to an additional screening and ultimately almost no gate action on the bottom TSS.
The same is true for screening by the bulk VB states in regimes III and IV \footnote{For explaining the reduced gate action, it is only important that bulk VB states start to be occupied at the boundary between regimes II and III and no distinction is made between regimes III and IV as in Ref.~\cite{Fuchs2025}.}.
Additionally, the gate action for both TSS is generally very small in regimes III and IV.
From the boundary between regimes II and III onward, toward lower densities, the chemical potential gets pinned to the large density of states of the VB (see Figs.~\ref{fig:dispersion_45nm}a, the Supporting Information, and Ref.~\cite{Mahler2021}).
While the total density still decreases with decreasing gate voltage, the position of the chemical potential is nearly constant, resulting in only small density changes for both TSS.
This ultimately prevents access to the Dirac point of HgTe 3DTIs in transport experiments \cite{Mahler2021,Fuchs2025}.
The smaller gate action for the bottom TSS in regimes I, III, and IV due to screening is also directly evident from the much smaller changes in the Hartree potentials towards the bottom of the well (cf. Figure~\ref{fig:potential_45nm}).
The very high transport mobilities of our samples (see appendix of Ref.~\cite{Fuchs2025}) allows resolving Shubnikov"~de~Haas oscillations from the bulk VB and CB states as well.
The calculated subband densities of the VB and CB are likewise in excellent agreement with their corresponding experimental counterparts.
The onset of the bulk VB states exactly coincides with the boundary between regimes II and III, while the CB states start to get occupied at the boundary between regimes I and II.
These results again confirm the boundaries and assignment of transport regimes I"~IV in Ref.~\cite{Fuchs2025}.

With our improved understanding of the carrier distribution in the sample, the choice of boundary conditions for the Hartree potential in the calculation arises naturally.
As the Rashba"~Bychkov"~type splitting of the TSS results from breaking of the structure inversion symmetry by the Hartree potential, no splitting occurs for symmetric Hartree potentials (cf.~Figure~\ref{fig:dispersion_45nm}b).
Hence, at the point where the density signatures from the top and bottom TSS are degenerate in the experiment, the resulting Hartree potential has to be symmetric.
With exception of the \SI{70}{nm} sample, this is the case for a total density of $\sim\SI{1.9e11}{cm^{-2}}$.
We discuss the origin of the finite density of the symmetric case and explain why the \SI{70}{nm} sample is symmetric at a smaller total density in the Supporting Information.
Using Gauss's law, the electric field boundary condition in the bottom barrier is chosen such that this density results in a symmetric Hartree potential.
A previous modeling attempt of thick HgTe quantum wells \cite{Gospodaric2020} did not consider the non"~trivial boundary conditions, which might explain some of the observed discrepancies.

\section{Conclusion}

By modeling the experimental subband density evolution in HgTe quantum wells with applied gate voltage, we have shown that the self-consistent Hartree method within the FB"~EFA gives numerically stable and quantitatively accurate results, even for very thick, topologically inverted layers, where the conventional approach fails.
We expect our openly-available implementation \cite{Beugeling2025} to greatly benefit the investigation of narrow-, broken-, and inverted"~gap materials, and facilitate applications such as quantum cascade lasers \cite{Nauschuetz2023}, electric field driven topological band inversion \cite{Li2009,Yang2008,Meyer2024}, efficient third"~harmonic THz generation \cite{UamanSvetikova2023,Svetikova2024}, and optoelectric modulator devices \cite{DaguaConda2025}.
Currently \kdotpy{} \cite{Beugeling2025} implements a zinc blende Hamiltonian but an expansion to other crystal structures is planned.

Here, we have considered a one-dimensional model at zero magnetic field for the electrostatics.
The excellent match between the self-consistent calculations and experimental data establishes the soundness of these approximations.
Nevertheless, for smaller devices or if the the gate electrodes cover the sides of the mesa, one may need to consider extra spatial dimensions.
At large magnetic fields, it might also be necessary to perform a self-consistent Landau level calculation.
Both cases come at a significantly higher computational expense, and we only expect minor corrections for the current investigation.
An effort to implement and test magnetic field and multi-dimensional self-consistent calculations using \kdotpy{} is already underway \cite{Beugeling2025}.

\FloatBarrier

\section*{Data Availability}
The band structure calculation software \kdotpy{} is available as an open"~source software project under \cite{Beugeling2025}.
The data underlying this study are openly available in Zenodo at \cite{}.

\section*{Supporting Information}
Symmetry considerations, variance in fitting parameters $\Delta\mu$ and $\epsilon_\text{HfO$_x$}$, origin of finite experimental intrinsic carrier density, Shubnikov"~de~Haas oscillations on reciprocal scale, analysis for extended thickness range (PDF)

\section*{Author Contributions}
M.H., C.B., and W.B. developed the self-consistent Hartree implementation in the band structure calculation software;
M.H. performed the calculations and analyzed the data with assistance from M.S. and M.S.;
C.F., L.F., and S.S. conducted the experiments;
H.B., T.K., W.B., and L.W.M. conceptualized the work;
M.H. and W.B. prepared the manuscript with input from all authors.

\section*{Acknowledgements}
We acknowledge financial support from the Deutsche Forschungsgemeinschaft (DFG, German Research Foundation) in the project SFB 1170 (Project ID 258499086) and in the Würzburg-Dresden Cluster of Excellence on Complexity and Topology in Quantum Matter \textit{ct.qmat} (EXC 2147, Project ID 390858490), and from the Free State of Bavaria (Institute for Topological Insulators).

\clearpage
\onecolumngrid

\begin{figure}[p]
	\centering
	\includegraphics[]{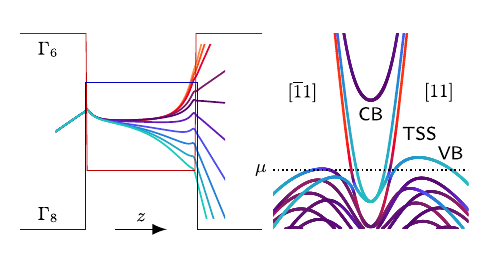}
	\caption{For Table of Contents Only.}
	\label{fig:toc}
\end{figure}

\clearpage

\begin{figure}[p]
	\centering
	\includegraphics[width=\linewidth]{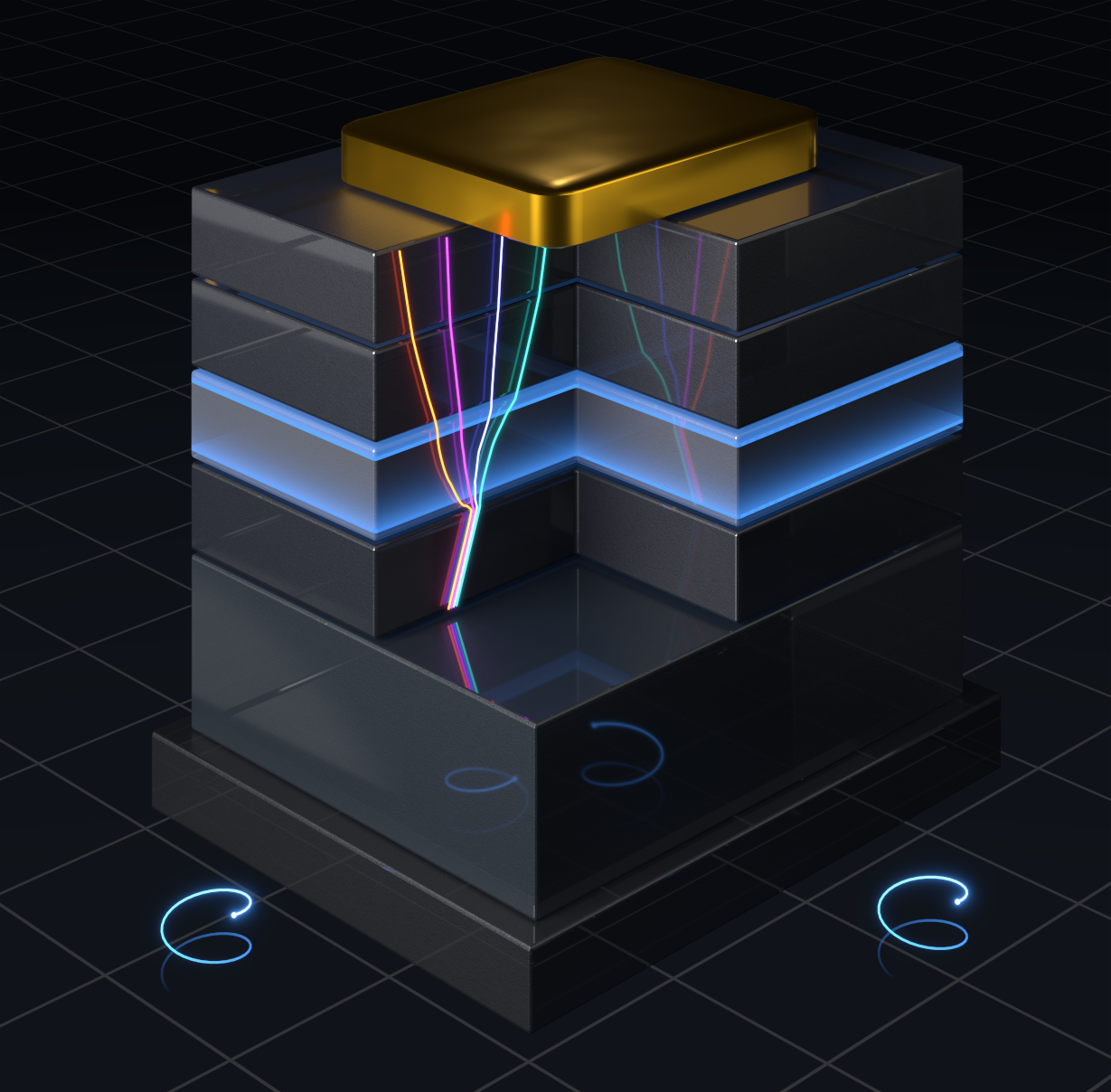}
	\caption{Cover art.
	A three-dimensional topological insulator (3DTI) heterostructure is subject to an external electric field from a top-side gate electrode.
	The resulting profound changes to the band structure, which lead to large Rashba-Bychkov splitting of the topological surface states, evidenced by Shubnikov-de Haas oscillations, are modeled using a self-consistent k.p Hartree approach.}
	\label{fig:cover_art}
\end{figure}

\clearpage
\appendix
\renewcommand{\thefigure}{S\arabic{figure}}
\setcounter{figure}{0}

\section*{Supporting Information}

\subsection*{Symmetry Considerations}

The combined effects of including non"~axial and bulk"~inversion asymmetry terms, uniaxial strain along the growth direction $z$, as well as breaking of the structure inversion symmetry ($z\to -z$) due to the Hartree potential, reduces the overall symmetry from $\overline{4}3m$ ($T_d$, bulk zinc blende) to only $mm2$ ($C_{2v}$).
The corresponding symmetry group in $(k_x,k_y)$"~space (i.e. in a 2D plane perpendicular to the $z$"~axis) is $2mm$ ($\mathrm{D}_2$).
This explains the splitting of the VB states into four disconnected hole pockets along the $[11]$/$[\overline{1} \overline{1}]$ (red) and $[\overline{1}1]$/$[1\overline{1}]$ (blue) directions, the red and blue pairs being equivalent by symmetry.
Due to the two"~fold valley degeneracy, only two distinct density signatures are expected in the experiment.
Hence, we calculate the density of the hole pockets along the $[11]$ and $[\overline{1}1]$ directions.

\begin{figure}[h]
    \centering
    \includegraphics[]{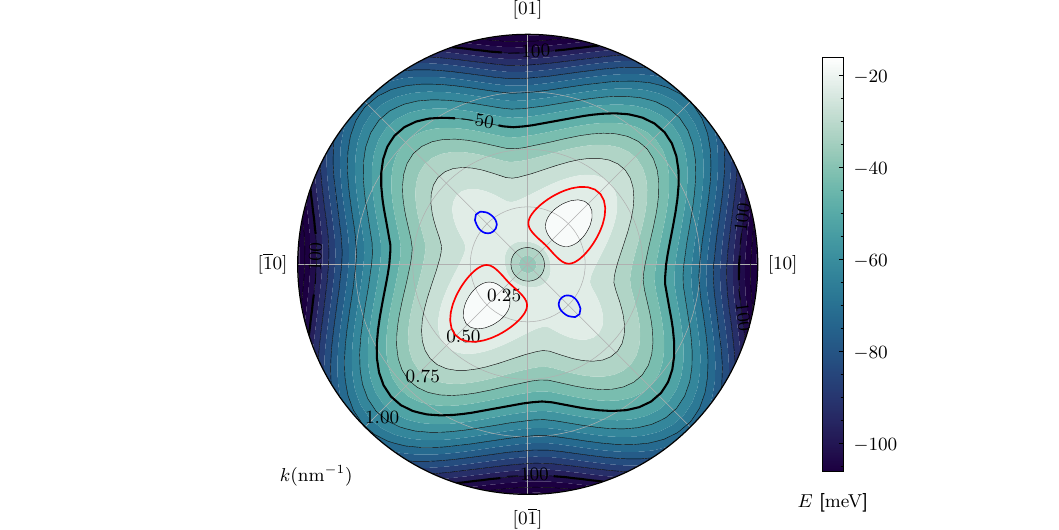}
    \caption{Energy dispersion $E$ of the uppermost bulk valence band of a \SI{45}{nm} thick HgTe quantum well at a total carrier density of $n_\text{tot}=\SI{-3.0e11}{cm^{-2}}$ for the corresponding self-consistent Hartree potential shown in Figure~\ref{fig:potential_45nm} in the main text.
    The position of the chemical potential is indicated by the red and blue lines, resulting in four disconnected hole pockets.
    The pairs along the $k$-directions $[11]$/$[\overline{1} \overline{1}]$ (red) and $[\overline{1}1]$/$[1\overline{1}]$ (blue) are equivalent by the $2mm$ symmetry.}
    \label{fig:VB_topography_45nm}
\end{figure}

\FloatBarrier
\subsection*{Variance in Fitting Parameters $\Delta\mu$ and $\epsilon_\text{HfO$_x$}$}

Across our four samples, the fitting procedure yields values between \SI{-130}{meV} and \SI{-280}{meV} for $\Delta\mu$ and between $5.2$ and $10$ for $\epsilon_\text{HfO$_x$}$.
The relatively large variance in $\Delta\mu$ likely arises from changes in the exact thickness and distribution of Ti, as the effective work function of the Ti/Au bilayer depends critically on the thickness of the Ti layer \cite{Lu2005}.
The fitted values for $\epsilon_\text{HfO$_x$}$ agree well with recent results ($\epsilon_\text{HfO$_x$}\sim 7.3$) on samples fabricated using the same low"~temperature ALD process for growing the HfO$_x$ layer \cite{Liang2024}.
Here, the large variance could be due to the considerable uncertainty (several \si{nm}) in the thickness of the top (Hg,Cd)Te barrier, which leads to uncertainty in the value of $U(z_\text{top})$.
The barrier thickness and $\epsilon_\text{HfO$_x$}$ both affect $V_g$ in the same way (see eq.~\ref{eq:VG}) and are thus statistically dependent.
Furthermore, the thickness of the gate dielectric $d_\text{HfO$_x$} \approx \SI{15}{nm}$ is also not precisely known and fluctuates across different samples.
For a well controlled and characterized sample fabrication process, $\Delta\mu$ and $\epsilon_\text{HfO$_x$}$ should only need to be fitted once, allowing for future predictions of gate voltages.

\subsection*{Origin of Finite Experimental Intrinsic Carrier Density}

The finite density and slightly negative gate voltage of the symmetric case might be explained by formation of trap states and interface dipoles at the (Hg,Cd)Te/HfO$_x$ interface during lithographic processing and/or the work function difference between the gate metal and quantum well, resulting in an effective gating from the top of initially almost charge neutral (prior to lithographic processing) samples \cite{Liang2024,Hinz2006,Lunczer2019}.
It was speculated that the formation of the same trap states also leads to a reduction of the inelastic scattering time \cite{Lunczer2019}.
The significantly smaller symmetric density of only $\sim\SI{0.9e11}{cm^{-2}}$ of the \SI{70}{nm} sample (cf.~Figure~\ref{fig:FFT_26nm_70nm_107nm}b) is consistent in this picture, as this sample also shows a more than two times larger peak transport mobility of $>\SI{2e6}{cm^2.V^{-1}.s^{-1}}$ compared to the other samples \cite{Fuchs2025}.

\FloatBarrier
\subsection*{Shubnikov"~de Haas Oscillations on Reciprocal Scale}

\begin{figure}[h]
    \centering
    \includegraphics[]{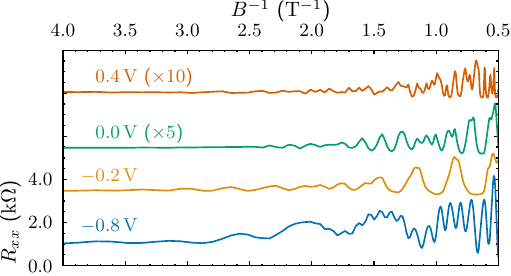}
    \caption{Experimental longitudinal resistance curves $R_{xx}$ in an out"~of"~plane magnetic field $B$ at different gate voltages of a \SI{45}{nm} thick HgTe quantum well.
    The absolute scale applies to the \SI{-0.8}{V} curve.
    For clarity, the other curves are offset vertically and the \SI{0.0}{V} and \SI{0.4}{V} curves are scaled as indicated.
    The curves are the same as already presented in Figure~\ref{fig:FFT_45nm}a in the main text but as a function of reciprocal magnetic field $1/B$.
    }
    \label{fig:FFT_45nm_sup}
\end{figure}

To emphasize the periodic nature of the Shubnikov"~de~Haas oscillations in $1/B$, Figure~\ref{fig:FFT_45nm_sup} presents the same measurements on the \SI{45}{nm} sample from Figure~\ref{fig:FFT_45nm}a in the main text on a reciprocal magnetic field scale.

\FloatBarrier
\subsection*{Analysis for Extended Thickness Range}

\begin{figure}[h]
  \centering
  \begin{tabular}{@{}c@{}}
    \includegraphics[]{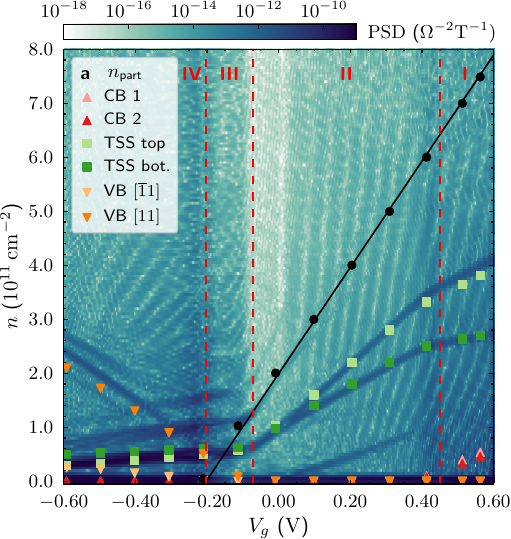} \\[\abovecaptionskip]
  \end{tabular}
  \begin{tabular}{@{}c@{}}
    \includegraphics[]{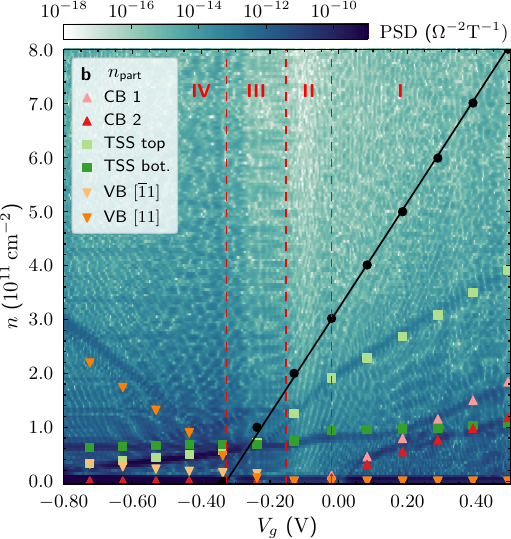} \\[\abovecaptionskip]
  \end{tabular}
  \begin{tabular}{@{}c@{}}
  \includegraphics[]{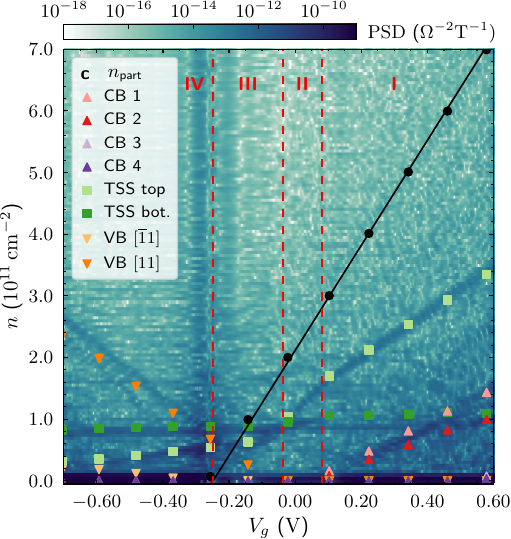} \\[\abovecaptionskip]
  \end{tabular}
  \caption{Subband densities of (a) a \SI{26}{nm}, (b) a \SI{70}{nm}, and (c) a \SI{107}{nm} thick HgTe quantum well.
  Colored symbols show the individual subband densities $n_\text{part}$ of both topological surface states (TSS), the first (CB 1,2) and second bulk conduction bands (CB 3,4), and the hole pockets at the top of the valence band (VB) from self"~consistent Hartree \kdotp{} calculations.
  The background shows the power spectral density (PSD) at density $n$ obtained from FFTs of the experimental low"~field Shubnikov"~de~Haas oscillations at different gate voltages $V_g$.
  The black line corresponds to the experimental gate action for the total carrier density obtained from Hall measurements.
  The black dots give the total carrier density from the calculations.
  Red"~dashed lines divide the gate voltage range into four transport regimes I-IV, see discussion in the main text and Ref.~\cite{Fuchs2025}.
  }
  \label{fig:FFT_26nm_70nm_107nm}
\end{figure}

Self"~consistently calculated subband densities and experimental FFT charts for additional \SI{26}{nm}, \SI{70}{nm}, and \SI{107}{nm} thick HgTe quantum wells are presented in Figure~\ref{fig:FFT_26nm_70nm_107nm}.
The subband density evolution is again reproduced quite well by the self"~consistent band structure calculations.
For the thinnest measured sample (\SI{26}{nm}, cf. Figure~\ref{fig:FFT_26nm_70nm_107nm}a), the splitting caused by the Hartree potential is least pronounced. 
Due to the large spatial wave function overlap between the top and bottom TSS, the screening strength is greatly reduced compared to the thicker quantum wells.
The large spatial wavefunction overlap might also give rise to magneto"~intersubband"~oscillations (MISO) between the TSS \cite{Minkov2022}.
This could explain the occurrence of the additional density signature in the experiment, corresponding to the density difference of both TSS, which starts around $V_g=\SI{0}{V}$ and grows in density for larger gate voltages (cf.~Figure~\ref{fig:FFT_26nm_70nm_107nm}a).
It is remarkable that even for such a thin sample the calculation still matches the experimental density signatures quite well.
This experimentally demonstrates the orthogonality of the two TSS, as else a strong scattering between the two states would occur that would smear out the individual density signatures, possibly to a point where only the total carrier density signature remains.

\end{document}